\documentclass[reprint,superscriptaddress,showkeys]{revtex4-1}
\pdfoutput=1
\usepackage[utf8]{inputenc}
\usepackage{amsmath}
\usepackage{amssymb}
\usepackage{amsfonts}
\usepackage{graphicx}
\usepackage{psfrag}
\usepackage{siunitx}
\usepackage{color}
\usepackage{soul}

\newcommand{\fluc}{\mathrm{fluc}}
\newcommand{\diss}{\mathrm{diss}}
\newcommand{\avg}{\mathrm{avg}}
\newcommand{\amb}{\mathrm{amb}}

\newcommand{\meas}{\mathrm{meas}}

\renewcommand{\d}{\mathrm{d}}

\begin{document}

\title{Extended equipartition in a mechanical system subject to a heat flow: the case of localised dissipation.}

\author{Alex Fontana}
\author{Richard Pedurand}
\author{Ludovic Bellon}
\email{Corresponding author : ludovic.bellon@ens-lyon.fr}
\affiliation{Univ Lyon, Ens de Lyon, Univ Claude Bernard Lyon 1, CNRS, Laboratoire de Physique, F-69342 Lyon, France}

\date{\today}

\begin{abstract}
Statistical physics in equilibrium grants us one of its most powerful tools: the equipartition principle. It states that the degrees of freedom of a mechanical system act as a thermometer: temperature is equal to the mean variance of their oscillations divided by their stiffness. However, when a non-equilibrium state is considered, this principle is no longer valid. In our experiment, we study the fluctuations of a micro-cantilever subject to a strong heat flow, which creates a highly non-uniform local temperature. We measure independently the temperature profile of the object and the temperature yielded from the mechanical thermometers, thus testing the validity of the equipartition principle out of equilibrium. We demonstrate how the fluctuations of the most energetic degrees of freedom are equivalent to the temperature at the base of the cantilever, even when the average temperature is several hundreds of degrees higher. We then present a model based on the localised mechanical dissipation in the system to account for our results, which correspond to mechanical losses localised at the clamping position.
\end{abstract}

\maketitle

\section{Introduction}

Thermal noise is the manifestation of random fluctuations of the microscopic constituents of any system having a non-zero temperature. In an experiment, its effect is a tiny variance around the mean value of the observable taken into consideration. Due to its intrinsic minute amplitude compared to other noise sources, the fluctuations usually go unnoticed. Nevertheless, their presence is significant in many contexts and must be considered. For example, microelectromechanical systems (MEMS) are systems that require a study of thermal noise, as it is the factor that limits their ultimate sensitivity \cite{Mohd2009}. In biology, cellular membranes present thermal fluctuations which must be quantified in order to understand the bioelectro-magnetism \cite{Vincze2005} and survival of cells \textit{in vitro} \cite{Johnson1972}. Gravitational waves (GW) detectors are limited in sensitivity by the thermal noise of the coating on the mirrors \cite{Harry2006}. Numerous other examples of technological applications or physical phenomena exist, as thermal fluctuations become salient when system size decreases or measurement sensitivity increases. Its understanding is thus fundamental.

When a system is in thermal equilibrium, the Fluctuation-Dissipation Theorem (FDT) \cite{Callen1951} constitutes an effective framework to study the fluctuations of an observable. In many applications, however, equilibrium is not a given. In living systems \cite{Gupta2017}, aging materials \cite{Buisson2004} and systems subject to a heat flux \cite{Monnet2019,Conti2013}, for example, the FDT cannot be expected to hold \textit{a priori}, and such aforementioned experiments are then necessary to test its validity beyond the thermodynamic equilibrium hypotheses. In many cases \textit{higher} fluctuations with respect to equilibrium are measured when out of equilibrium, thus violating the FDT~\cite{Conti2013,Cugliandolo2011,Lumbroso2018}. The opposite behavior has also been observed: the flexural degrees of freedom of a silicon micro-cantilever strongly out of equilibrium present a \textit{lower} noise compared to equilibrium \cite{Geitner2016}. In this experiment, the cantilever is a spatially-extended system brought into a non equilibrium steady state (NESS) by a strong heat flux: a laser heats one side of the system while the other side is kept at room temperature. Thermal fluctuations are measured, and the results show a deficit of fluctuations: there is almost no difference between the thermal noise in an equilibrium situation at room temperature and a NESS one, when the average temperature is several hundreds degrees higher. This peculiar behavior is explained thanks to an extension of the FDT that arises from the inspirational work of Levin \cite{Levin1998} relating the fluctuations of the flexural normal modes (i.e. the degrees of freedom) to a localized mechanical energy dissipation of the cantilever.

In this work, we demonstrate that these results can be extended to the torsional degrees of freedom of a similar cantilever, thus completing the aforementioned study. We consider the same system under a strong heat flux and measure the simultaneous thermal fluctuations of flexural and torsional modes at once, showing that they are indeed insensitive to the average temperature rise. We then generalize the FDT for the torsional degrees of freedom to account for our results. Our study could be useful to understand out of equilibrium systems, to guide investigations of thermodynamics far from equilibrium, and to help engineer low-noise instruments.

In the first section, we present the experimental setup, showing how we create a NESS and measure thermal noise. In the following sections, we show the outcome of the experiment and develop the theory to account for it. The final section discusses the results.

\section{Methods}

\begin{figure}
\begin{center}
\includegraphics[width=0.5\textwidth]{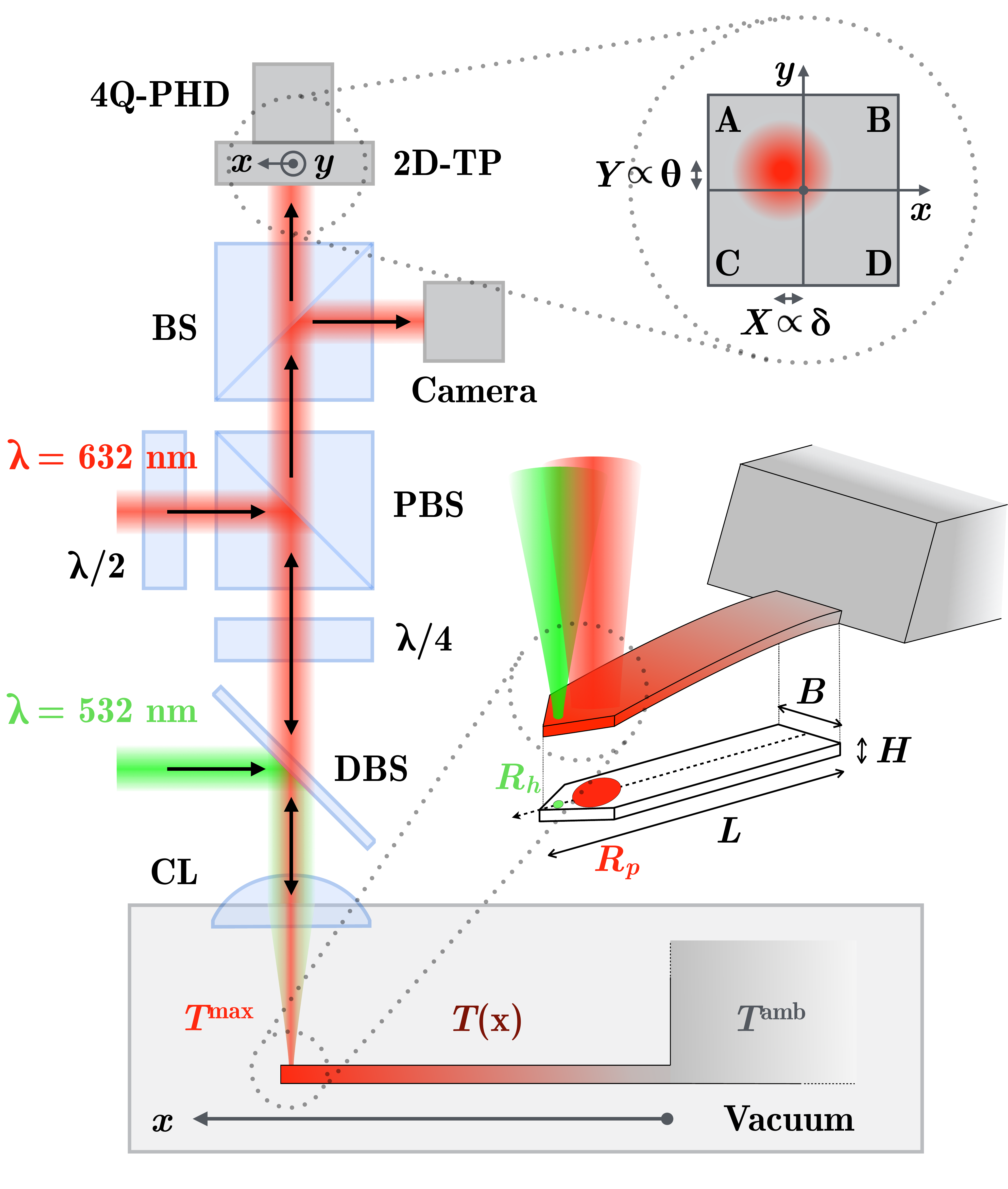}
\caption{Experiment setup: the deflection and torsion of a cantilever are captured thanks to the optical lever technique. A red laser beam ($\SI{1}{mW}$ at $\SI{633}{nm}$) probes the deformations close to the tip of the cantilever. It enters the system through a half-wave plate ($\lambda/2$) tuning its polarisation so that after passing through the polarising beam splitter (PBS), all light is directed towards the cantilever. It then passes through a quarter-wave plate ($\lambda/4$), a dichroic beam splitter (DBS), and a converging lens (CL, focal length $f_{CL}=\SI{30}{mm}$) which focuses the beam on the cantilever tip. The lens is also used as the light port to the vacuum chamber. Light is reflected back on the same path from the cantilever. The second passage through the $\lambda/4$ rotates the polarisation perpendicular to the initial one, therefore the return beam passes straight through the PBS. A final beam splitter (BS) divides it towards an optical camera, used to position the lasers on the cantilever, and the four quadrants photodiode (4Q-PHD). This latter sensor records the temporal signals of deflection $\delta(t)$ and torsion $\theta(t)$ as the difference of intensity in the quadrants in the $x$ and $y$ directions respectively. A motorised 2D translation platform (2D-TP) controlling the position of 4Q-PHD in these directions is used during the calibration procedure (see Appendix \ref{appendix:cal}). The absorption of a green laser beam ($0$ to $\SI{10}{mW}$ at $\SI{532}{nm}$) focused close to the tip of the triangular end of the cantilever acts as the heater. It is directed towards the cantilever by the DBS and through the lens. The reflected light runs through the same path out of the system. The two lasers spots do not overlap in order to avoid mutual disturbances. The cantilever, in vacuum at $\SI{5e-6}{mbar}$, is monolithically clamped to its macroscopic chip which is thermalised at room temperature $T^\amb$.}
\label{Path}
\end{center}
\end{figure}

\begin{figure}[t]
\begin{center}
\includegraphics[width=0.5\textwidth]{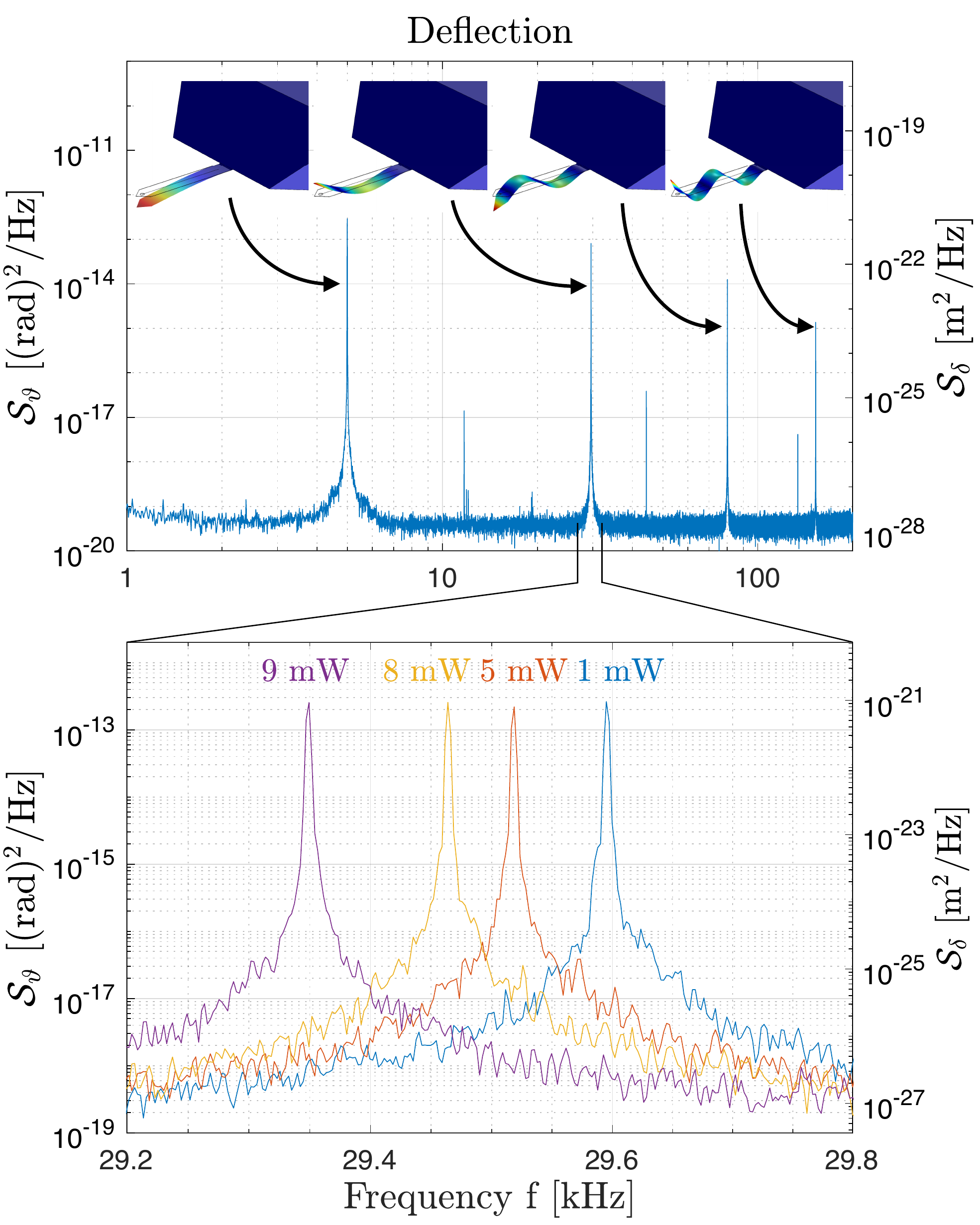}
\caption{PSDs of the thermal noise-induced deflection of the cantilever. In the upper plot, each resonance mode is identified as a sharp peak with a quality factor in the range of tens of thousands. The modes can safely be considered decoupled and each can be treated as a simple harmonic oscillator. In the lower figure, a zoom-in around the second flexural resonance shows how the resonance is redshifted with increasing laser power. This phenomenon is used to compute the imposed $\Delta$T. The shapes of the modes are simulated in COMSOL \cite{COMSOL}, yielding resonance frequencies very close to the ones found in our experiment and in good agreement with the Euler-Bernoulli description. The left axis of the plots corresponds to the measured flexural angles $\vartheta$ by the optical lever detection (in $\SI{}{rad^2/Hz}$), while the right axis corresponds to its conversion for deflection $\delta$ (in $\SI{}{m^2/Hz}$, using mode 2 sensitivity).}
\label{Spectrad}
\end{center}
\end{figure}

\begin{figure}[t]
\begin{center}
\includegraphics[width=0.5\textwidth]{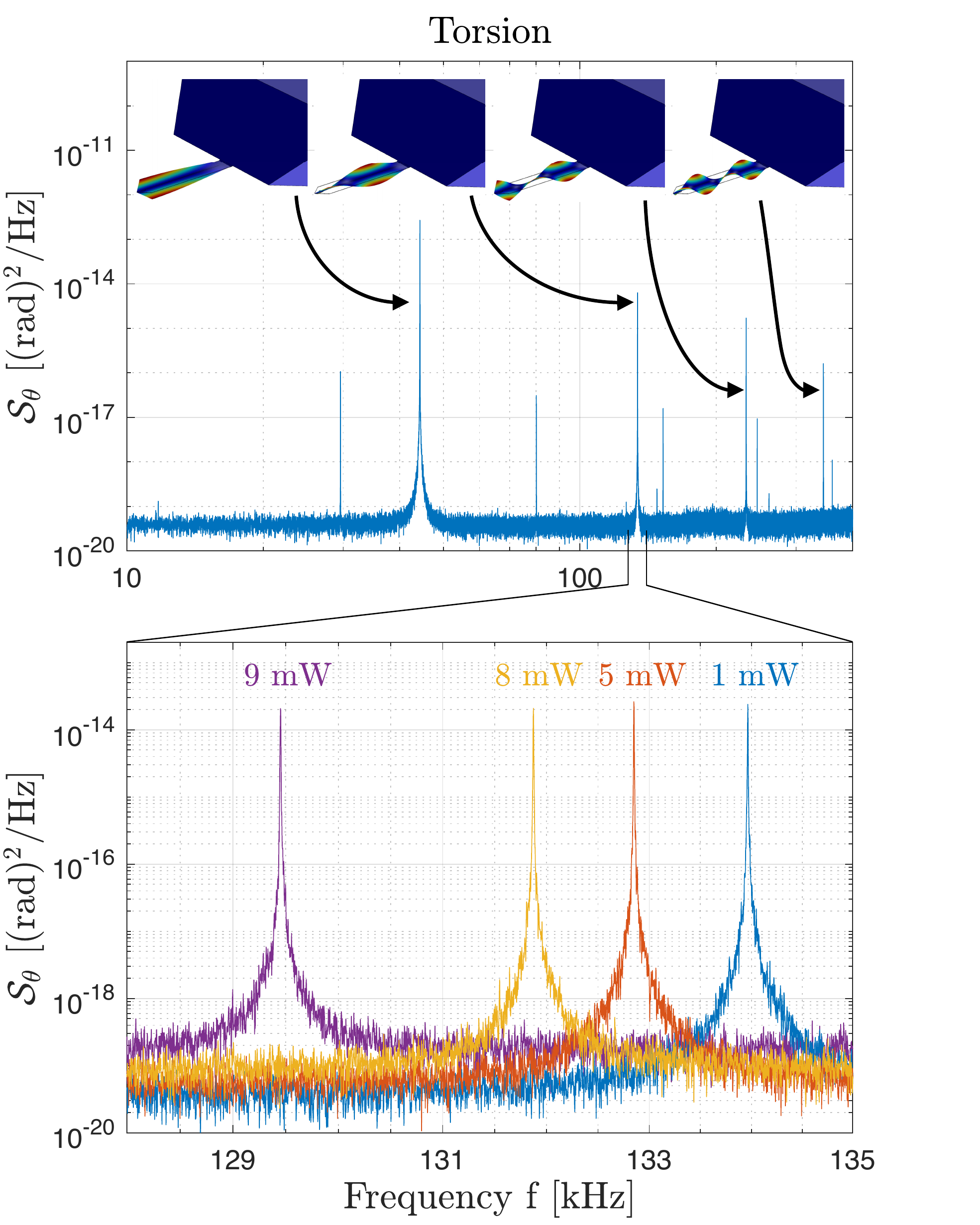}
\caption{PSDs of the thermal noise-induced torsion of the cantilever. In the upper plot each resonance mode is identified as a sharp peak with a quality factor in the range of tens of thousands. In the lower figure, a zoom-in around the second torsional resonance shows how the resonance is redshifted with the laser power increasing, comparatively more with respect to deflection modes. With the model currently at hand, $\Delta$T cannot be calculated with enough precision through torsional frequency shift. The simulated frequencies of the resonances agree quite accurately with the experiment, whereas for the higher modes the analytical Saint-Venant model deviates from the observation.}
\label{Spectrat}
\end{center}
\end{figure}

As pictured in fig.~\ref{Path}, the physical system considered is a $L=\SI{500}{\mu m}$ long, $B=\SI{100}{\mu m}$ wide and $H=\SI{1}{\mu m}$ thick silicon cantilever (Nanoworld Arrow TL-8) \cite{ARROW} monolithically clamped to a macroscopic chip. It is placed in a vacuum chamber at $\SI{5e-6}{mbar}$. Thermal noise is measured close to the free end of the cantilever thanks to the optical lever technique \cite{Jones1961,Meyer1988}: a $\SI{633}{nm}$ laser is focused at normal incidence on the cantilever and its reflection is collected with a four quadrant photodiode. Processing the signals along the $y$-axis of the photodetector leads to the calibrated torsional angle $\theta$ (in rad), while the $x$-axis leads to the calibrated flexural angle $\vartheta$ (in rad), which can be converted to the deflection $\delta$ (in m, see Appendix \ref{appendix:cal} for details). The waist diameter is tuned to roughly $2R_p=\SI{100}{\mu m}$ to maximize sensitivity \cite{Gustafsson1994}. Computing the Power Spectrum Density (PSD), we identify the normal modes of the cantilever (see fig.~\ref{Spectrad} and \ref{Spectrat}). The spectra are shot noise limited and the thermal noise-driven resonance peaks show a high signal to noise ratio. The resonances have very high quality factors, usually tens of thousands, therefore they can be identified as independent degrees of freedom, each behaving as a simple harmonic oscillator \cite{Butt1995}. Typical measurements are $\SI{150}{s}$ data sets sampled at $\SI{2.5}{MHz}$, allowing us to explore a wide range of frequencies including up to 11 flexural and 8 torsional modes. Due to experimental constraints, in this work we focus on the flexural modes spanning from 2 to 8 and on the torsional ones from 1 to 8, excluding mode 5 in both cases (see Appendix \ref{appendix:cal} for details).

\subsection{Heating}

The cantilever is heated by a second $\SI{532}{nm}$ laser, focused near the free triangular end of the cantilever. The waist diameter is tuned to $2R_h=\SI{10}{\mu m}$, and the spots of the two lasers do not overlap (gap of around $\SI{10}{\mu m}$). The base of the cantilever is monolithically clamped to its macroscopic silicon chip, which acts as a thermal reservoir at room temperature $T^\amb$. In vacuum, the most efficient way to dissipate the heat is through conduction, thus a temperature difference $\Delta T$ is established along the cantilever length. The characteristic time for heat diffusion in the cantilever is $\SI{2.5}{ms}$ at room temperature, so thanks to the constant laser power pouring energy into the tip, we can safely assume that $\Delta T$ is stationary. Therefore, the system can be regarded as in a steady state. In these conditions, a huge temperature difference can be reached with just a few mW of laser power: at roughly $\SI{9}{mW}$ the temperature at the tip $T^{\max}$ is around 700 K higher than the temperature at the base (see fig.~\ref{Teffs}).

The details for the temperature gradient estimation can be found in \cite{Aguilar2015}; we summarise here the procedure. When heated, the cantilever experiences a change in its stiffness due to the evolution of its Young modulus and to thermal expansion. As it turns softer for high $T$, the cantilever's normal oscillation modes occur at a smaller angular frequencies $\omega$ (fig.~\ref{Spectrad} and \ref{Spectrat}, bottom panels). $\Delta T$ can be extracted by tracking the frequency shift with respect to the equilibrium situation, when no heating is present. The consequent temperature profile $T(x)$ of the cantilever can be calculated, allowing us to define its mean temperature $T^\avg$. Here $x$ is the dimensionless longitudinal coordinate of the cantilever. 

\subsection{Thermal noise}

Even if the system can be described in a local thermal equilibrium framework (meaning a local temperature $T(x)$ can always be defined), the thermal noise of each mode corresponds to a collective motion of the whole cantilever. In equilibrium, fluctuations of an observable (in our case, $\delta$ or $\theta$) are characterized by a single temperature $T$ thanks to the Equipartition Principle (EP): 
\begin{equation}
\label{eq:EP}
k \langle \delta^2\rangle = \kappa \langle \theta^2 \rangle = k_B T 
\end{equation}
with $\langle \delta^2 \rangle$, $\langle \theta^2 \rangle$ the mean square deflection and torsion, calculated as an integration of the PSD in a tiny band around the resonances, $k$, $\kappa$ the respective stiffnesses, $k_B$ the Boltzmann constant. Once the system is in a NESS, this relation does not necessarily hold anymore, and a universal temperature cannot be easily defined. Still, the cantilever oscillates under the action of its thermal fluctuations following the motion of each normal modes motion, thus every resonance carries the information of the thermal energy content. Since the resonances are well separated and uncoupled, this results in as many thermometers as the number of resonances at hand, possibly showing different values. This system is therefore a well-suited test bench for systems in a NESS, being small enough to have uncoupled many degrees of freedom as a thermal noise probes and large enough to allow strong temperature gradients. We will thus write:
\begin{equation}
\begin{split}
\label{eq:EPNESS}
k_{n} \langle \delta^2_n \rangle & = k_B T^{\fluc}_n \\
\kappa_{m} \langle \theta^2_m \rangle & = k_B T^{\fluc}_m 
\end{split}
\end{equation}
where the quantities depend now from the mode number $n,m$ for the deflection and torsion respectively. In addition, $T$ is now called \emph{fluctuation temperature} as it represents a temperature in an out of equilibrium system. In this regime, in fact no thermodynamic temperature of the cantilever can be defined, and the $T^{\fluc}_{n,m}$ embody the meaningful value of the fluctuation amplitudes. 

\section{Results}

\begin{figure}
\begin{minipage}{\columnwidth}
\includegraphics[width=\textwidth]{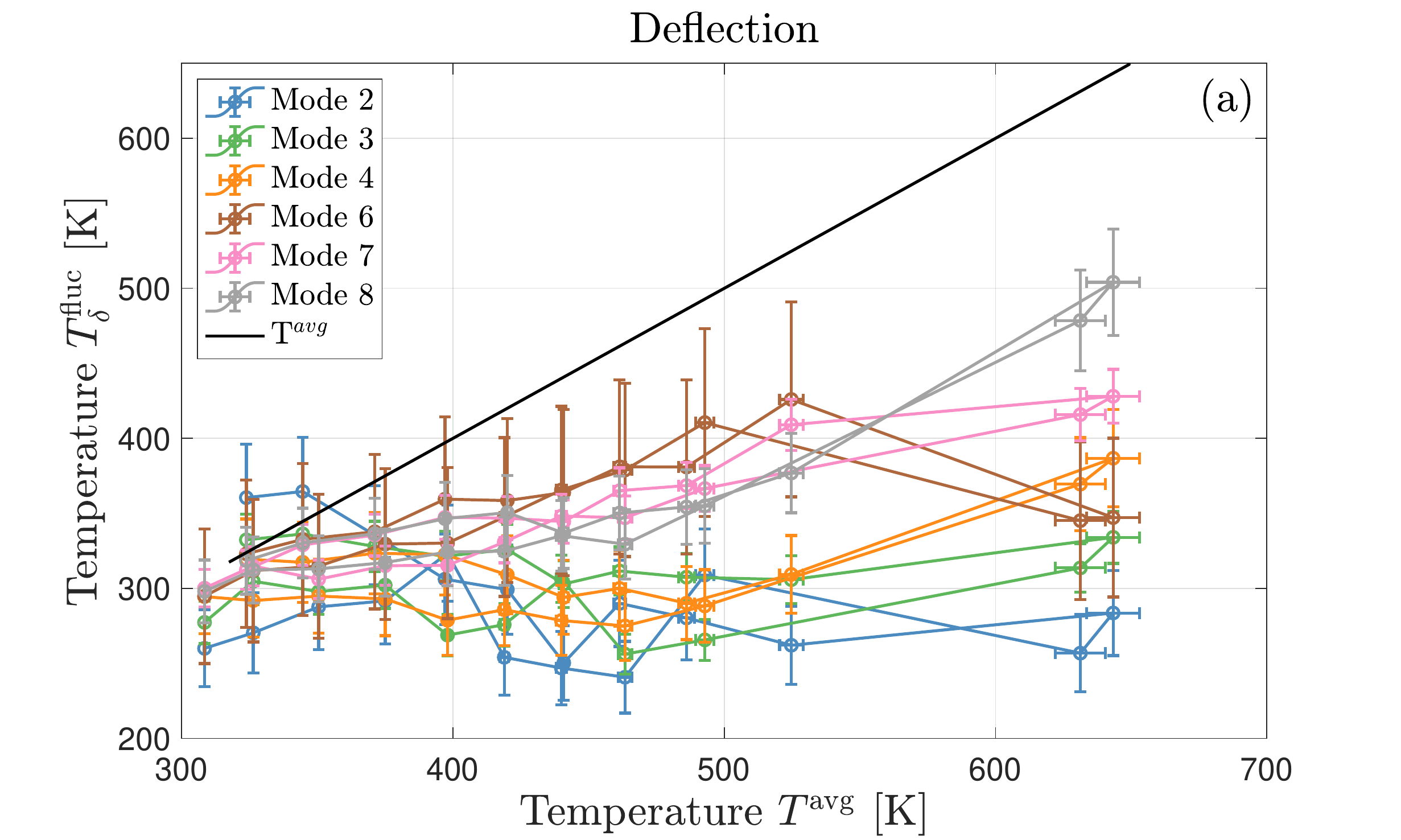}
\end{minipage}
\begin{minipage}{\columnwidth}
\includegraphics[width=\textwidth]{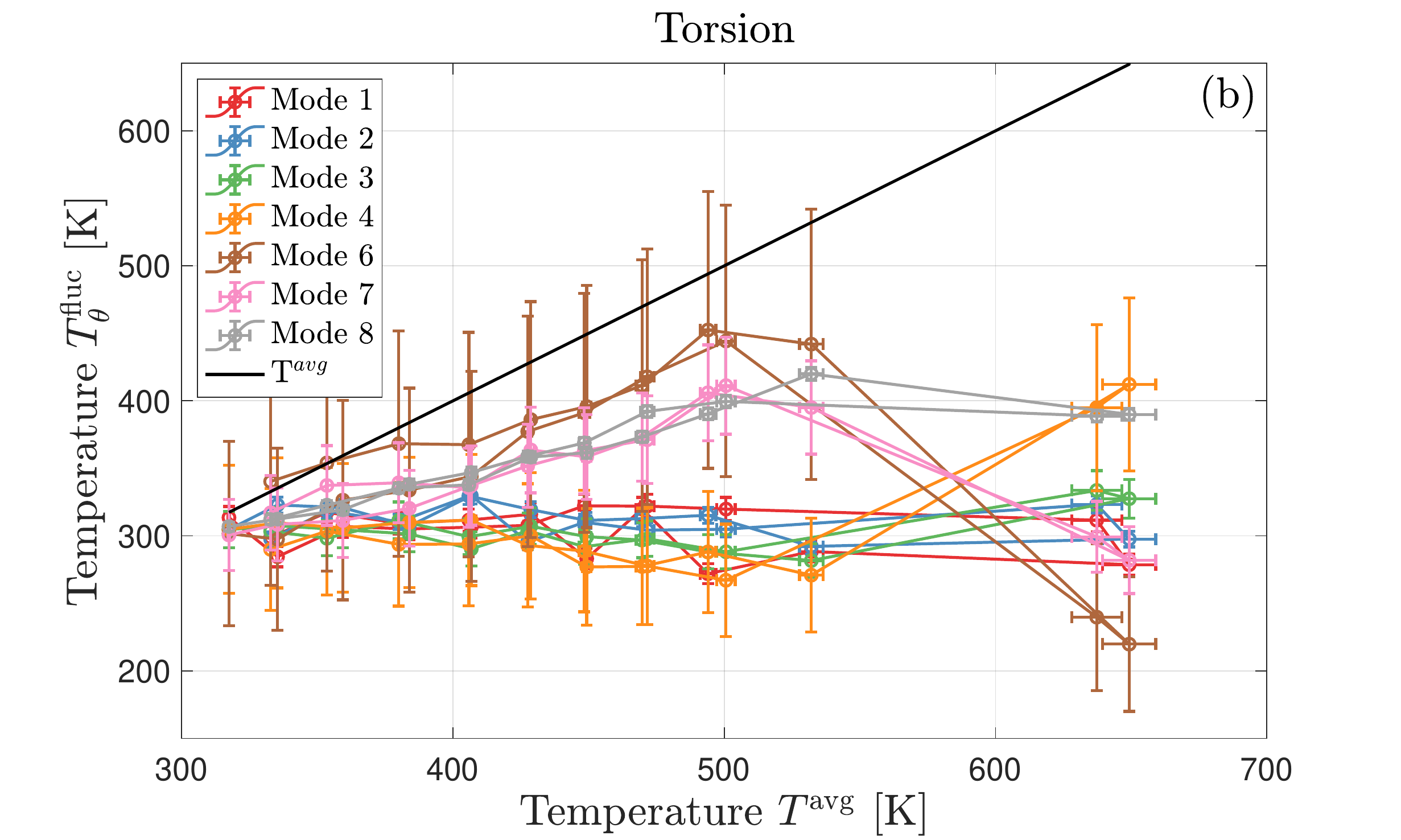}
\end{minipage}
\caption{Fluctuation temperature vs. average temperature. In figure (a), the flexural $T^{\fluc}_n$ is shown with respect to $T^\avg$. The black line represents the``equilibrium" temperature, i.e. the fluctuations an object would show had it been in thermal equilibrium with a thermal bath at $T^\avg$. All the modes lie below this line, as if there was a dearth of thermal noise. As explained in appendix \ref{appendix:ruling}, the results for the first deflection mode were strongly disturbed by a parasitic self oscillation phenomenon, and are therefore not presented. The modes shown span from 2 to 8, excluding mode 5 because of a lack of sensitivity at the laser probe position. In figure (b), the same scenario is shown for the torsional degrees of freedom (and mode 5 is omitted for the same reason). See Appendix \ref{appendix:Tfluc} for details in computing $T^{\fluc}$.}
\label{Teffs}
\end{figure}

We present in fig.~\ref{Teffs} the measured fluctuation temperatures for both flexural and torsional motions as a function of the average temperature of the system. 

In the run, 10 different laser powers are set, thus allowing us to probe a maximum temperature $T^{\max}$ going from 330 K up to 1000 K and then backwards. This two-sided ramp is meant as a test of robustness for our method and to assure that we don't alter the material during the measurement. The flexural fluctuation temperature is roughly constant whilst the temperature of the cantilever increases, thus confirming the results in \cite{Geitner2016}. We find in fact the same behavior, and extend the previous study for more resonances: the fluctuations do not appear to change sensibly with the system going out of equilibrium. This tendency is highlighted by the points lying \emph{below} the average temperature curve, thus showing a \emph{deficit} of fluctuations. The torsional fluctuation temperatures further asses this phenomenon: the thermal noise is unaffected by the temperature rise in the system for this observable as well. Looking at COMSOL simulations, we verify that we have probed all the existing modes in the explored frequency range, showing that they all present lower fluctuations than equilibrium.

Let us now present a theoretical approach accounting for these results.

\subsection{First approach}

These observations were construed in \cite{Geitner2016} for the flexural modes, showing how a careful extension of the FDT for this out of equilibrium system leads to an expression of the fluctuation temperature:
\begin{equation}
\label{eq:TEFF}
T^{\fluc}_n = \int_0^1 dx T(x) w^\diss_n(x)
\end{equation}
where $w^\diss_n$ is the normalised mechanical energy dissipation density for mode $n$, that acts as a weight on the temperature profile. The integral is along the normalized length of the cantilever, with $x\equiv x/L$. In ref.~\cite{Geitner2016} it is shown that the fluctuation temperatures are unchanged whilst the average temperature increases for a cantilever similar to the one of the current study. As such, the dissipation has to be located at the base of the cantilever, so that the only relevant temperature is $T^\amb$. As a matter of fact, the cantilever is etched from a single cristal silicon wafer, i.e. in principle devoid of internal defects \footnote{The sole contribution onto internal damping due to thermoelastic yields a quality factor of the order of $10^6-10^7$, at least $10$ times higher than what is measured \cite{Lifshitz2000}}, and the vacuum removes most of the hydrodynamical damping \footnote{This statement is even more true in our experiment than in \cite{Geitner2016}, since the pressure is lowered by a factor $\sim10^4$}. The only part that can present some defects is thus the clamped end, where the chemically etched cantilever is attached at the chip, resulting in $T^{\fluc}_n \approx T^\amb$. Our experiment yields compatible results with this description, and so we believe that our system is similarly  characterized by a local dissipation occurring at the base. Therefore, We therefore confirm the validity of our model to describe deflection modes up to 8 and it appears that it can be extended to the torsional degrees of freedom. A careful calculation is presented in the next section. 

\subsection{Current model}

In his remarkable approach, Levin \cite{Levin1998} demonstrates that in equilibrium, if an observable is a weighted function of the deformation, its thermal noise PSD is proportional to the energy dissipation of the system submitted to a force distributed according to the same spatial distribution. We measure the deflection (or torsion) of a cantilever, which can be decomposed on the base of its normal modes as:
\begin{equation}
\delta(x,t) = \sum_{n=1}^{\infty} \delta_n(t) \phi_n(x)
\end{equation}
The quality factor of each resonance mode is very high, thus around each resonance the PSD of $\delta$ can actually be identified with that of the corresponding mode amplitude $ \delta_n$. Our observables are therefore simply the deformation of the cantilever weighted by the mode shape:
\begin{equation}
\delta_n(t) = \int_{0}^{1} dx \delta(x,t) \phi_n(x)
\end{equation}
Following ref. \cite{Levin1998}, thermal noise associated with $ \delta_n$ at equilibrium is then:
\begin{equation}
\label{eq:SdtL}
\mathcal{S}_{\delta_n}(\omega) = \frac{4 k_B}{\pi \omega^2} T \int dx \frac{W^\diss_{n}(x,w)}{F_n^2}
\end{equation}
where $F_n$ is the amplitude of the force $F(x,t)=F_n \phi_n(x) \cos(\omega t)$ giving rise to the average dissipated power $W^\diss_{n}(x,\omega)$. 

To compute the mean square deflection $\langle \delta_n ^2 \rangle$ of mode $n$, we need to integrate the PSD over all frequencies. Since the resonance is very sharp, integrating in a narrow frequency range around it leads to the same result:
\begin{align}
\langle \delta_n ^2 \rangle &= \int_0^\infty d\omega \mathcal{S}_{\delta_n}(\omega) \label{eq:Teffstep1}\\
\frac{k_B T^\fluc_n}{k_n} &= \int_0^\infty d\omega \frac{4 k_B}{\pi \omega^2} T \int_0^1 dx \frac{W^\diss_{n}(x,\omega)}{ F_n^2}\\
T^\fluc_n &= T \int_0^1 dx \int_0^\infty d\omega \frac{4 k_n W^\diss_{n}(x,\omega)}{\pi \omega^2 F_n^2}\\
T^\fluc_n &= T \int_0^1 dx \, w^\diss_{n}(x) \label{eq:Teffstep4}
\end{align}
where
\begin{equation} \label{eq:wdissn}
w^\diss_{n}(x) = \int_0^\infty d\omega \frac{4 k_n W^\diss_{n}(x,\omega)}{\pi \omega^2 F_n^2}
\end{equation}
is proportional to the dissipated power of mode $n$. Since in equilibrium $T^\fluc_n = T$, we see that the integral over $x$ of $w^\diss_{n}(x)$ is 1, thus $w^\diss_{n}(x)$ is the normalised dissipated power of mode $n$.

Komori and co-workers \cite{Komori2018} extend the work of Levin to NESSes presenting a distribution of temperature rather than an equilibrium temperature. Their result is very similar to Levin's, except that in eq.~\ref{eq:SdtL}, the temperature field $T(x)$ is included under the integral:
\begin{equation}
\label{eq:SdtW}
\mathcal{S}_{\delta_n} (\omega)= \frac{4 k_B}{\pi \omega^2}\int dx T(x) \frac{W^\diss_{n}(x,\omega)}{F_n^2}
\end{equation}
Repeating the steps from eqs. \ref{eq:Teffstep1} to \ref{eq:Teffstep4}, we immediately retrieve the expression of $T^\fluc_n$ given by eq.~\ref{eq:TEFF}, with $w^\diss_{n}(x)$ proportional to the dissipated power of mode $n$ and still defined by eq.~\ref{eq:wdissn}. The only remaining criterion to verify is that this quantity is still normalised (in the sense that its integral over all $x$ is 1) when the system is out of equilibrium. Let us first note that any effect of the temperature gradient on $w^\diss_{n}(x)$ will be a second order effect for $T^\fluc_n$, which is already proportional to $T(x)$. Moreover, the mechanical response of the cantilever is only slightly modified by the temperature gradient, since the maximum frequency shift (and so the stiffness) registered is at most a few percent. To a very good approximation, $w^\diss_{n}(x)$ can thus be considered as the normalised dissipation of mode $n$. In appendix \ref{Appendix:visco}, an explicit formula is given for $w^\diss_{n}(x)$ when the energy dissipation can be described by the loss tangent of the material.

The previous demonstration has been conducted for the flexural modes, but it applies just as well to torsion. The deformation of the cantilever can be decomposed on the normal modes in torsion $\phi_m(x)$, and the thermal noise of the each mode amplitude $\theta_m$ is given by:
\begin{equation}
\label{eq:Sthetan}
\mathcal{S}_{\theta_m}(\omega) = \frac{4 k_B}{\pi \omega^2}\int dx T(x) \frac{W^\diss_{m}(x,\omega)}{\Gamma_m^2}
\end{equation}
where $\Gamma_m$  is the amplitude of the torque $\Gamma(x,t)=\Gamma_m \phi_m(x) \cos(\omega t)$ giving rise to the dissipated power $W^\diss_{m}(x,\omega)$. Integrating over frequencies to compute the mean square torsion $\langle \theta_m ^2 \rangle$ of mode $m$, we find again
\begin{equation}
\label{eq:TEFFm}
T^{\fluc}_m = \int_0^1 dx T(x) w^\diss_m(x)
\end{equation}
where $w^\diss_{m}(x)$ is the normalised dissipation of mode $m$ in torsion.

Eqs. \ref{eq:TEFF} and \ref{eq:TEFFm} for deflection and torsion are equivalent to the expression of ref. \cite{Geitner2016}, though they are derived from a different approach. Moreover, the current equations apply to torsion as well. This model is thus anticipated to describe our experimental observations: the fluctuation temperatures of both families of modes correspond to the average of the temperature field weighted by the local mechanical energy dissipation.

\section{Discussion}
When the system is in thermal equilibrium, one recovers eq.~\ref{eq:EP} from the previous equations, since in this case simply $T^{\fluc}=T$. When the system is in a NESS, the amplitude of fluctuations quantified by $T^{\fluc}$ is proportional to the temperature profile weighed by the mechanical energy dissipation. In our experiments, we show that thermal noise for both deflection and torsion is unaffected by the temperature rise in the tip, thus confirming the results in \cite{Geitner2016} and showing that this is true for the first (and most energetic) resonances of the cantilever. Consequently, We can hereby affirm that the cantilevers considered in these experiments show a \emph{lack} of fluctuations as a general feature.

The model presented in order to asses these observation follows the latest studies \cite{Komori2018} around NESS systems, showing how we can construct the non-equilibrium PSDs of the cantilever thermal noise through the dissipated mechanical energy. The importance of our experiment is thus twofold: 1) it makes use of the aforementioned technique and shows that it can be applied to multiple resonances of a system, and 2) it validates the approach itself demonstrating that it leads to a deep physical meaning. The simple procedure of plugging the temperature $T(x)$ into the dissipation integral in the work of Komori (eq.~1) leads in our case to a satisfactory understanding of our experimental data. As a result, we believe that this procedure can have a broad range of applications, owning to the generality of the underlying method. Possibly noise in nano-mechanical resonators \cite{Cleland2002} can be predicted when the system is subject to a temperature gradient; gravitational waves detection in cryogenic conditions is now at the testing phase \cite{Akutsu2019}, and the out of equilibrium state of the suspended mirrors has to be taken into account \cite{Komori2018}; Johnson noise can be modeled through the proposed description when a $\Delta T$ is applied \cite{Monnet2019}. Furthermore, experiments such as the aforementioned are necessary to test the validity of the latest theoretical predictions over fluctuation theorems and the relative inequalities \cite{Horowitz2020}. 

To conclude, this work shows a way to naturally extend the FDT in the case of a NESS. Fluctuations should lie between the minimal and maximal temperature of the system, and our experiment shows the lower extreme case. In other systems as the aforementioned ones, the opposite situation may also arise, showing an increase of fluctuations up to the highest temperature, and possibly beyond~\cite{Conti2013}. Universality is not yet (and may never be) attainable when considering non equilibrium thermodynamics, thus experiments and theoretical development is encouraged. Future work may include measuring thermal noise in different situations: changing the oscillator (material, shape) and boundary conditions (pressure, clamp temperature). 

\section*{Acknowledgments}
We would like to thank A. Petrosyan for technical support and J. Pereda for stimulating discussions.

\appendix

\section{Calibration of the measurement} \label{appendix:cal}

\subsection{Photodetector}

Upon reflexion on the cantilever end, the laser beam is deviated according to the slope of the cantilever at the focal point of the lens CL. Let $\vartheta_\meas$ be the angle corresponding to the deflection, and $\theta_\meas$ be the angle corresponding to torsion: the light is deflected by twice those angles. Following the lens of focal length $f_{CL}$, the collimated laser beam will thus be shifted by $X=2 f_{CL} \vartheta_\meas$ and $Y=2 f_{CL} \theta_\meas$, which correspond to the coordinates of the center of the laser spot on the photodetector (PHD-4Q in fig.~\ref{Path}). The four quadrants of this split photodiode record four signals, namely $A, B, C, D$ (top left, top right, bottom left, and bottom right, respectively), from which we evaluate two contrasts :
\begin{equation}
\begin{split}
\label{eq:d&t}
C_x &= \frac{(A+C)-(B+D)}{A+B+C+D} \\
C_y &= \frac{(A+B)-(C+D)}{A+B+C+D}
\end{split}
\end{equation}
These are dimensionless numbers, proportional to the spot position $(X,Y)$ on the photodetector for small displacements. For a gaussian beam of $1/e^2$ radius $R_x$ for example, one has $C_x = X/R_x$ for $X\ll R_x$. A simple calibration procedure to determine $R_x$ is to use the 2D translation platform housing the photodetector: with the cantilever still, we shift the sensor origin and record $C_x$ for a few values of $X$ around 0, then perform a linear fit to directly extract $R_x$ from the slope. Note that since $R_x$ is extracted from a measurement, the beam shape can deviate from gaussianity with no influence on our results. The same calibration can be performed in the $y$ direction to measure $R_y$. The incertitude of these coefficients is typically $\le 0.1 \%$.  Eventually, the calibrated measurements of the slopes are given by:
\begin{align}
\vartheta_\meas & = \frac{R_x C_x}{2f_{CL}} \\
\theta_\meas & = \frac{R_y C_y}{2f_{CL}}
\end{align}

Using a calibrated motorised 2D translation platform, this procedure is automated and performed at each heating power. Indeed, the change in temperature affects the calibration, changing the static curvature of the cantilever for example. Between no heating and the most intense one, the relative difference between the calibration factors is around $10\%$. This would result in a $20\%$ difference in terms of fluctuation temperature if this calibration procedure was not performed.

\subsection{Laser spot position influence}

The reading of the photodetector thus leads to the knowledge of the slopes of the cantilever at the laser spot position $x_0$. We are rather interested in the amplitudes of all modes at the end of the cantilever: vertical deflection $\delta_n$ (units: m) and angular torsion $\theta_m$ (units: rad). For each mode, we can define a sensitivity as:
\begin{align}
\vartheta_{\meas, n} & = \sigma_n^\delta (x_0) \delta_n \\
\theta_{\meas, m} & = \sigma_m^\theta (x_0) \theta_m
\end{align}
For a small spot size, the sensitivity is simply linked to the normal mode shapes:
\begin{align}
\sigma_n^\delta(x_0)&=\frac{1}{\phi_n(L)}\frac{\d\phi_n}{\d x} (x_0) \label{eq:sigma_n} \\
\sigma_m^\theta(x_0)&=\frac{1}{\phi_m(L)}\phi_m (x_0) \label{eq:sigma_m}
\end{align}
However, since the spot size is large (around $\SI{100}{\mu m}$ in diameter), these relations only hold for low mode numbers: for large $n$ and $m$, the slope is not constant on the lighted area, and a more accurate sensitivity has to be computed. Following Schäffer \cite{Schaffer2005}, one can show that:
\begin{align}
\sigma_n^\delta(x_0) & \propto \int_0^L \d x \int_0^L \d\widetilde{x} E(x,x_0) E(\widetilde{x} ,x_0) \frac{\phi_n(x)-\phi_n(\widetilde{x})}{x-\widetilde{x}} \label{eq:fullsigma_n} \\
\sigma_m^\theta(x_0) & \propto \int_0^L \d x E(x,x_0)^2 \phi_m(x) \label{eq:fullsigma_m}
\end{align}
with $E(x,x_0) \propto e^{-(x-x_0)^2/R_p^2}$ the field of the gaussian beam of $1/e^2$ radius $R_p$. The missing proportionality factor can be computed using the constraint that eqs. \ref{eq:sigma_n} and \ref{eq:sigma_m} are recovered in the limit $R_p \rightarrow 0$.

To allow for a quantitative comparison between measurements, it is therefore important to keep the probe laser beam position $x_0$ constant during the full experiment. From the camera view, we can estimate the position of the probe beam to $x_0=\SI{400}{\mu m}$, and that the maximum drift during one experiment is limited to $dx_0=\SI{3}{\mu m}$ at most. This position is chosen to accommodate for the triangular tip and beam diameter, and it corresponds to a zero in sensitivity for mode 5 in deflection and in torsion: $\sigma_5^\delta(x_0)\approx0$ and $\sigma_5^\theta(x_0)\approx0$.

\subsection{Mean square values of the fluctuations}

Finally, the calibrated mean square values of the deflection and torsion for each mode are computed by integrating their PDS in a narrow frequency range $2 \Delta f \approx \SI{6}{kHz}$ around each resonance. Starting from the signals $C_x$ and $C_y$ available from the photodetectors and including the calibration coefficients, we compute:
\begin{equation} \label{eq:fullcal}
\begin{split}
\langle \delta^2_n \rangle & = \left(\frac{R_x}{2 f_{CL} \sigma_n^\delta(x_0)}\right)^2 \int_{f_n \pm \Delta f} \mathcal{S}_{C_x}(f) \\
\langle \theta^2_m \rangle & = \left(\frac{R_y}{2 f_{CL} \sigma_m^\theta(x_0)}\right)^2 \int_{f_m \pm \Delta f} \mathcal{S}_{C_y}(f)
\end{split}
\end{equation}

\section{Computing $T^\fluc$ and uncertainties} \label{appendix:Tfluc}

An experiment typically consists of 100 measurements, $\SI{2}{s}$ long, made for each laser power, and 10 laser powers are used in a two-sided ramp, thus resulting in roughly 2000 temporal signals. A calibration of the photodetector is performed after each step in laser power. From the photodiode signals, we first compute $C_x$ et $C_y$ (eqs. \ref{eq:d&t}), then their PSDs $\mathcal{S}_{C_x}$ and $\mathcal{S}_{C_y}$, with the Welch method \cite{Welch1967} using a Hann window \cite{Blackman1958} and an overlap of 50$\%$. Finally, we compute $\langle \delta^2_n \rangle $ and $\langle \theta^2_m \rangle$ with eqs. \ref{eq:fullcal}, for which we thus have 100 estimations at each laser power. In the following paragraphs we explain how we compute $T^\fluc$ and the statistical and systematic uncertainty from these 2000 estimations.

\subsection{$T^\fluc$ and statistical uncertainty}
The fluctuation temperatures in fig.~\ref{Teffs} are calculated as follows: $\langle \delta^2_n \rangle$, $\langle \theta^2_m \rangle$ of all datasets corresponding to the same power $P$ are averaged to estimate the mean square deflection and torsion at this specific power. A linear fit of those values versus $P$, for $P < 4$ mW, is computed and the ordinates at the origin $\langle \delta^2_n \rangle_0$, $\langle \theta^2_m \rangle_0$ are taken as the mean square fluctuations at equilibrium. We identify this value with $T^\amb$ through the equilibrium EP (eq.~\ref{eq:EP}) and subsequently we normalise the thermal noise values by it. Multiplying those ratio by $T^\amb=300 K$ yields the self-calibrated $T^{\fluc}$:
\begin{equation} \label{eq:Tfluctnorm}
\begin{split} 
T^{\fluc}_n &= \frac{\langle \delta^2_n \rangle}{\langle \delta^2_n \rangle_0} T^\amb\\
T^{\fluc}_m &= \frac{\langle \theta^2_n \rangle}{\langle \theta^2_n \rangle_0} T^\amb
\end{split}
\end{equation}
Note that this strategy is valid as long as great care is taken to ensure that the amplitude of fluctuations is comparable between all datasets, i.e. that $R_x$ and $R_y$ are measured, and $x_0$ is constant during all the experiment.

The uncertainties reported in fig. \ref{Teffs} are evaluated by the quadratic sum of two terms: a statistical contribution $\sigma^{stat}$ and a systematical one $\sigma^{sys}$. The former is evaluated thanks to the repeated measurement at each power, i.e. $\sigma^{stat}$ is the standard deviation around the mean. The latter is discussed below. 

\subsection{Systematic uncertainties}

The systematic uncertainties are related to the uncertainty $dx_0$ on the position $x_0$  of the laser probe on the cantilever, which may drift with time, for example with thermal expansion. From the camera view and thermal expansion computation, we estimate that $dx_0=\SI{3}{\mu m}$ at worst. As seen in eq. \ref{eq:Tfluctnorm} and \ref{eq:fullcal}, $T^{\fluc}_{n} \propto 1/\sigma_n^\delta(x_0)^2$, thus
\begin{equation}
\frac{1}{T^{\fluc}_n }\frac{\partial{T^{\fluc}_n}}{\partial x_0} = -2 \frac{1}{\sigma_n} \frac{\d\sigma_n^\delta}{\d x_0}
\end{equation}
Hence we calculate the systematic error as:
\begin{equation} \label{eq:sysuncertainty}
\sigma_n^{sys} = 2 \left|\frac{T^{\fluc}_n }{\sigma_n^\delta} \frac{\d\sigma_n^\delta}{\d x_0}\right| dx_0
\end{equation}
The same apply to torsion by replacing $\sigma_n^\delta(x_0)$ by $\sigma_m^\theta(x_0)$. Both these functions can be evaluated thanks to eqs. \ref{eq:fullsigma_n} and \ref{eq:fullsigma_m}. Table \ref{table:sysuncertainty} reports the estimated systematic uncertainties evaluated by this method. The sensitivity of mode 5 for both deflection and torsion being very close to 0, uncertainty is huge and those measurements are not presented. The same applies to flexural mode above mode 8, which present too high uncertainties.

\begin{table}[htbp]
\caption{Systematic uncertainties computed from eq. \ref{eq:sysuncertainty} for deflection modes 1 to 10 and torsion modes 1 to 8.}
\begin{center}
\begin{tabular}{||l||c|c|c|c|c|c|c|c|c|c||}
\hline
Mode number & 1 & 2 & 3 & 4 & 5 & 6 & 7 & 8 & 9 & 10 \\
\hline
Deflection: $\sigma_n^{sys}\,[\SI{}{K}$] & 5 & 2 & 10 & 48 & 643 & 92 & 21 & 40 & 258 & 231 \\
\hline
Torsion: $\sigma_m^{sys}\, [\SI{}{K}$] & 4 & 4 & 27 & 95 & 977 & 136 & 51 & 1 & - & - \\
\hline
\end{tabular}
\end{center}
\label{table:sysuncertainty}
\end{table}

\section{Explicit formula for the dissipation for viscoelastic damping} \label{Appendix:visco}
\subsection{Deflection}
We consider the Euler-Bernoulli framework to describe the deflection $\delta(x,t)$, or equivalently $\delta(x,\omega)$ in the Fourier space. The equation of motion (EOM) writes: 
\begin{equation}
\label{eq:EOMEB}
\biggl[- m \omega^2  + \frac{\partial^2}{\partial x^2}\biggl(k(x,\omega)\frac{\partial^2}{\partial x^2}\biggr)\biggr] \delta(x,\omega)=F(x,\omega)
\end{equation}
where $m$ is the mass of the cantilever, $k$ its stiffness (proportional to the Young's modulus) and $F$ an external driving force. In the most general case, $k$  can depend on frequency, but this dependency is always slow and can be forgotten around any specific resonance by replacing $\omega$ with $\omega_n$. The normal modes $\phi_n$ are the eigenvectors of the spatial operator in the EOM verifying the boundary conditions in absence of forcing, specifically:
\begin{equation}
\label{eq:normalmodes}
\frac{\partial^2}{\partial x^2}\biggl(k(x,\omega_n)\frac{\partial^2}{\partial x^2}\biggr) \phi_n(x) = k_n \phi_n(x)
\end{equation}
with $k_n$ the mode equivalent stiffness.

When the energy dissipation in the cantilever is mostly due to internal friction, sometimes referred to as viscoelasticiy \cite{Nowick1972}, it is usually expressed as an imaginary part in the elastic modulus of the material embedded in the stiffness:
\begin{equation}
\label{eq:visc}
k = k^0(1 + i \varphi_k)
\end{equation}
where $\varphi_{k}$ is the loss angle of the material for the considered deformation (both parameters $k^0$ and $\varphi_k$ can depend on position and frequency). In this case $W^\diss$ takes the form:
\begin{equation} \label{eq:Wdissphi}
W^\diss(x,\omega) = \omega \varphi(x,\omega) U^{\max}(x,\omega)
\end{equation}
where $U^{\max}$ is the energy of the cantilever when maximally strained: 
\begin{equation}
\label{eq:Upot}
U^{\max}_{\delta}(x,\omega) = \frac{1}{2} k^0(x,\omega)\biggl|\frac{\partial^2 \delta}{\partial x^2}\biggr|^2
\end{equation}
which can be easily derived from the EOM (eq.~\ref{eq:EOMEB}). Since we consider only the normal mode $n$ to compute $w^\diss_{n}(x)$, we can express the deformation as $\delta(x,\omega)=\delta_n(\omega) \phi_n(x)$. From the definition of $w^\diss$ by eq.~\ref{eq:wdissn}, we deduce
\begin{equation}
w^\diss_{n}(x) = \int_0^\infty d\omega \frac{2 k_n}{\pi \omega}\left|\frac{ \delta_n(\omega)}{F_n}\right|^2  \varphi_k(x,\omega) k^0(x,\omega)\phi_n^{\prime \prime 2} (x)
\end{equation}
with $\phi_n^{\prime \prime 2} (x)$ the double spatial derivative of the mode shape. In the integral over all frequencies, it should be noted that the term $|\delta_n/F_n|^2$ is the square of the frequency response of an harmonic oscillator with a large quality factor. Since it is thus highly peaked at the resonance frequency $\omega_n$ of the mode, all the parameters that are slowly varying functions of the frequency can be replaced by their values at $\omega_n$. We thus get:
\begin{equation}
w^\diss_{n}(x) = \varphi_k(x,\omega_n) k^0(x,\omega_n)\phi_n^{\prime \prime 2} (x) \int_0^\infty d\omega \frac{2 k_n \delta_n(\omega)^2}{\pi \omega |F_n|^2}  
\end{equation}
All the terms dependent on the position are out of the integral, which is therefore just a multiplicative factor. It can be evaluated from the response of the harmonic oscillator, or more simply by noting that $w^\diss_{n}(x)$ is normalized: $\int_0^1 dx w^\diss_{n}(x) = 1$, hence
\begin{equation}
w^\diss_{n}(x) = \frac{\varphi_k(x,\omega_n) k^0(x,\omega_n)\phi_n^{\prime \prime 2} (x)}{\int_0^1 dx \varphi_k(x,\omega_n) k^0(x,\omega_n)\phi_n^{\prime \prime 2} (x)}  
\end{equation}
If the dissipation is localised at the origin, then we can express it as a Dirac's delta centered in this point $\delta_D(x)$: $\varphi_k(x,\omega_n) \propto \delta_D(x)$, hence $w^\diss_{n} = \delta_D(x)$. This situation corresponds to the experimental data presented in this article.

\subsection{Torsion}

The framework considered for torsion is the Saint-Venant one~\cite{Landau1970}. The EOM now writes: 
\begin{equation}
\label{eq:EOMStV}
\biggl[- I \omega^2  + \frac{\partial}{\partial x}\biggl(\kappa(x,\omega)\frac{\partial}{\partial x}\biggr)\biggr] \theta(x,\omega)=\Gamma(x,\omega) 
\end{equation}
where $I=mB^2/12$ is the second moment of inertia of the cantilever, $\kappa$ its local torsional stiffness (proportional to the shear modulus) and $\Gamma$ an external torque. The normal modes $\phi_m$ are the eigenvectors of the spatial operators in the EOM verifying the boundary conditions in absence of forcing, specifically:
\begin{equation}
\frac{\partial}{\partial x}\biggl(\kappa(x,\omega_m)\frac{\partial}{\partial x}\biggr) \phi_m(x) = \kappa_m \phi_m(x)
\end{equation}
with $\kappa_m$ the equivalent stiffness of the mode.

When the energy dissipation in the cantilever is viscoelastic, we can write just as for the deflection:
\begin{equation}
\kappa = \kappa^0(1 + i \varphi_\kappa)
\end{equation}
with $\varphi_\kappa$ the loss angle. $W^\diss$ takes the same form as for deflection in eq.~\ref{eq:Wdissphi}, with 
\begin{equation}
U^{\max}_{\theta}(x,\omega) = \frac{1}{2} \kappa^0(x,\omega) \biggl|\frac{\partial \theta}{\partial x}\biggr| ^2
\end{equation}
Since we consider only the normal modes $m$ to compute $w^\diss_{m}(x)$, we decompose the deformation as $\theta(x,\omega) = \theta_m(\omega) \phi_m(x)$.

From the definition of $w^\diss$ by eq.~\ref{eq:wdissn}, we deduce
\begin{equation}
w^\diss_{m}(x) = \int_0^\infty df \frac{2 \kappa_m}{\pi \omega} \left|\frac{\theta_m(\omega)}{\Gamma_m}\right|^2  \varphi_\kappa(x,\omega) \kappa^0(x,\omega)\phi_m^{\prime 2} (x)
\end{equation}
Again, the square of the frequency response of the harmonic oscillator in torsion $|\theta_m/\Gamma_m|^2$ is highly peaked at the resonance frequency $\omega_m$ of the mode, so we replace all other parameters by their values at $\omega_m$. Using the normalisation property of $w^\diss_{m}(x)$, we eventually get
\begin{equation}
w^\diss_{m}(x) = \frac{\varphi_\kappa(x,\omega_m) \kappa^0(x,\omega_n)\phi_m^{\prime 2} (x)}{\int_0^1 dx \varphi_\kappa(x,\omega_m) \kappa^0(x,\omega_m)\phi_m^{\prime 2} (x)}  
\end{equation}
If the dissipation is located at the origin, then $\varphi_\kappa(x,\omega_m) \propto \delta_D(x)$, hence $w^\diss_{m} = \delta_D(x)$. This situation corresponds to the experimental data presented in this article.

\section{Ruling out external noise contributions}\label{appendix:ruling}

Thermal noise is usually minute and thus a difficult quantity to measure. The system under study has to be very small to have measurable thermal fluctuations, therefore exposing it to the possible interference of other perturbations that, however small, are usually orders of magnitude higher than the phenomenon we are interested in. In our experiment we hence took deep care of excluding any kind of external perturbations, isolating the system from the noise of the environment with a air suspended optical table, removing acoustic contributions and hydrodynamic interactions by placing the cantilever in vacuum, and securing that the laser does not interfere with thermal noise, adding or subtracting energy to the modes. 

In the next sections, we evaluate the various noises that can spoil the thermal noise estimation.

\subsection{Background electronic noise contribution}\label{appendix:Back}

The measured signal, for example $C_x$, is the sum of the actual thermal noise contribution $\delta$ and an electronic background noise contribution $N$:
\begin{equation}
C_x = S \delta + N
\end{equation}
with $S$ the sensitivity of the measurement.
The thermal noise is thus evaluated by
\begin{equation}
\langle \delta^2 \rangle = \frac{1}{S^2} (\langle C_x^2 \rangle - \langle N^2 \rangle)
\end{equation}
The last term is mainly due to the shot-noise of the photodiodes. As seen on fig. \ref{Spectrad}, it has a white noise behavior and is usually many order of magnitude below the resonances, hence it has a very low impact on the final result. It is nevertheless subtracted after measuring it off resonance, since its magnitude is independent on frequency. In order to ensure that the electronic background is indeed negligible, a second method is used: a cross-correlation technique \cite{Pottier2017}. Two distinct flexural contrast are calculated as:
\begin{equation}
\begin{split}
\label{eq:cdd}
C_{x1} &= \frac{A-B}{A+B} \\
C_{x2} &= \frac{C-D}{C+D}
\end{split}
\end{equation}
Supposing that the signal is the sum of thermal noise $\delta$ and shot-noise contribution $N_{1,2}$, we have $C_{x1,2} = S \delta+ N_{1,2}$. Computing the cross correlation between $C_{x1}$ and $C_{x2}$ leads to: 
\begin{align}
\label{eq:cdd2}
\langle C_{x1}C_{x2} \rangle &= S^2 \langle \delta^{2} \rangle + S \langle \delta N_{1}\rangle + S \langle \delta N_{2}\rangle + \langle N_{1}N_{2} \rangle\\
&= S^2 \langle \delta^{2} \rangle
\end{align}
where all but the first contribution are zero due to the noises being uncorrelated. Note that the same strategy applies to torsion by changing the pairs of quadrant to compute $C_{y1}=(A-C)/(A+C)$ and  $C_{y2}=(B-D)/(B+D)$. For all the modes the difference between the methods is less than 1$\%$, therefore the electronic background noise has little if no influence on the results. 

\subsection{Laser power fluctuations}

Due to radiation pressure and photo thermal effects, any fluctuation of the laser power can translate into a force on the cantilever, thus a fluctuation in deflection or torsion possibly biasing the results. We thus estimate the sensitivity of the measured mean square deflection $\langle \delta^{2} \rangle$ (and torsion $\langle \theta^{2} \rangle$) due to the laser fluctuations. A white noise is added to the laser power thanks to the Acousto-Optic Modulator (AOM) \cite{AOM} that is routinely used as the power controller, and the signals of incoming laser power $P_\mathrm{driven}$ alongside $\delta_\mathrm{driven}$ are measured. The gain of the transfer function $\chi_{P,\delta}$ is computed as the ratio between the PSDs of the two:
\begin{equation}
|\chi_{P,\delta}| = \mathcal{S}_{\delta_\mathrm{driven}}/\mathcal{S}_{P_\mathrm{driven}}
\end{equation}
Once this transfer function has been characterised, a measurement without any additional noise is performed and the new PSD of laser power is multiplied by the $|\chi_{P,\delta}|$. The ratio between $|\chi_{P,\delta}|\mathcal{S}_{P}$ and $\mathcal{S}_{\delta}$ represent the amount of laser driven fluctuations. For the first two flexural and the first torsional mode, for all laser powers this contribution is less than $1\%$:
\begin{align}
|\chi_{P,\delta}\mathcal{S}_{P}| &\ll \mathcal{S}_{\delta}\\
|\chi_{P,\theta}\mathcal{S}_{P}| &\ll \mathcal{S}_{\theta}
\end{align}
For higher order modes, the transfer function is even weaker and hard to characterize, so the same conclusions apply.

\subsection{Self oscillations of first deflection mode}
As a final remark, the first flexural mode is excluded from the results because of an optomechanical coupling between the laser heating the system and the cantilever \cite{Metzger2008}. This phenomenon is different from the laser pollution discussed above and causes a strong increase or decrease of fluctuations during most of the measurement time, therefore hiding the thermal noise. We decide to avoid presenting its $T^{\fluc}$ due to the poor statistics it has once the corrupted points are eliminated.

\bibliographystyle{ieeetr}
\bibliography{ExtEP-locdiss}

\begin{thebibliography}{10}

\bibitem{Mohd2009}
F.~Mohd-Yasin, D.~J. Nagel, and C.~E. Korman, ``Noise in {MEMS},'' {\em
  Measurement Science and Technology}, vol.~{21}, p.~012001, nov 2009.

\bibitem{Vincze2005}
G.~Vincze, N.~Szasz, and A.~Szasz, ``On the thermal noise limit of cellular
  membranes,'' {\em Bioelectromagnetics}, vol.~{26}, pp.~28--35, 01 2005.

\bibitem{Johnson1972}
H.~J. Johnson and M.~Pavelec, ``Thermal noise in cells. a cause of spontaneous
  loss of cell function,'' {\em Am J Pathol.}, vol.~{69}, pp.~119--130, 10
  1972.

\bibitem{Harry2006}
G.~M. Harry, H.~Armandula, E.~Black, D.~R.~M. Crooks, G.~Cagnoli, J.~Hough,
  P.~Murray, S.~Reid, S.~Rowan, P.~Sneddon, M.~M. Fejer, R.~Route, and S.~D.
  Penn, ``Thermal noise from optical coatings in gravitational wave
  detectors,'' {\em Applied Optics}, vol.~{45}, 7, 2006.

\bibitem{Callen1951}
H.~B. Callen and T.~A. Welton, ``Irreversibility and generalized noise,'' {\em
  Phys. Rev.}, vol.~{83}, pp.~34--40, Jul 1951.

\bibitem{Gupta2017}
S.~K. Gupta and M.~Guo, ``Equilibrium and out-of-equilibrium mechanics of
  living mammalian cytoplasm,'' {\em Journal of the Mechanics and Physics of
  Solids}, vol.~{107}, pp.~284 -- 293, 2017.

\bibitem{Buisson2004}
L.~Buisson, M.~Ciccotti, L.~Bellon, and S.~Ciliberto, ``{Electrical noise
  properties in aging materials},'' in {\em Fluctuations and Noise in
  Materials} (D.~Popovic, M.~B. Weissman, and Z.~A. Racz, eds.), vol.~{5469},
  pp.~150 -- 164, International Society for Optics and Photonics, SPIE, 2004.

\bibitem{Monnet2019}
B.~Monnet, S.~Ciliberto, and L.~Bellon, ``Extended nyquist formula for a
  resistance subject to a heat flow,'' {\em Journal of Statistical Mechanics:
  Theory and Experiment}, vol.~2019, no.~10, p.~104011, 2019.

\bibitem{Conti2013}
L.~Conti, P.~D. Gregorio, G.~Karapetyan, C.~Lazzaro, M.~Pegoraro, M.~Bonaldi,
  and L.~Rondoni, ``Effects of breaking vibrational energy equipartition on
  measurements of temperature in macroscopic oscillators subject to heat
  flux,'' {\em Journal of Statistical Mechanics: Theory and Experiment 2013},
  vol.~{12}, P12003, 2013.

\bibitem{Cugliandolo2011}
L.~F. Cugliandolo, ``The effective temperature,'' {\em Journal of Physics A:
  Mathematical and Theoretical}, vol.~{44}, p.~483001, nov 2011.

\bibitem{Lumbroso2018}
O.~S. Lumbroso, L.~Simine, A.~Nitzan, D.~Segal, and O.~Tal, ``Electronic noise
  due to temperature differences in atomic-scale junctions,'' {\em Nature},
  vol.~562, no.~7726, pp.~240--244, 2018.

\bibitem{Geitner2016}
M.~Geitner, F.~Aguilar~Sandoval, E.~Bertin, and L.~Bellon, ``Low thermal
  fluctuations in a system heated out of equilibrium,'' {\em Physical Review
  E}, vol.~{95}, 12 2016.

\bibitem{Levin1998}
Y.~Levin, ``Internal thermal noise in the ligo test masses: A direct
  approach,'' {\em Phys. Rev. D}, vol.~{57}, pp.~659--663, Jan 1998.

\bibitem{ARROW}
Nanoworld. \url{https://www.nanoworld.com/array-of-8-}.

\bibitem{Jones1961}
R.~V. Jones, ``Some developments and applications of the optical lever,'' {\em
  Journal of Scientific Instruments}, vol.~{38}, pp.~37--45, feb 1961.

\bibitem{Meyer1988}
G.~Meyer and N.~M. Amer, ``Novel optical approach to atomic force microscopy,''
  {\em Applied Physics Letters}, vol.~{53}, no.~12, pp.~1045--1047, 1988.

\bibitem{Gustafsson1994}
M.~G.~L. Gustafsson and J.~Clarke, ``Scanning force microscope springs
  optimized for optical beam deflection and with tips made by controlled
  fracture,'' {\em Journal of Applied Physics}, vol.~{76}, no.~1, pp.~172--181,
  1994.

\bibitem{Butt1995}
H.~J. Butt and M.~Jaschke, ``Calculation of thermal noise in atomic force
  microscopy,'' {\em Nanotechnology}, vol.~{6}, pp.~1--7, jan 1995.

\bibitem{COMSOL}
S.~COMSOL~AB, Stockholm, ``Comsol
  multiphysics\textsuperscript{\textregistered}.''
  \url{https://www.comsol.com}.
\newblock v. 5.4.

\bibitem{Aguilar2015}
F.~Aguilar~Sandoval, M.~Geitner, E.~Bertin, and L.~Bellon, ``Resonance
  frequency shift of strongly heated micro-cantilevers,'' {\em Journal of
  Applied Physics}, vol.~{117}, 03 2015.

\bibitem{Note1}
The sole contribution onto internal damping due to thermoelastic yields a
  quality factor of the order of $10^6-10^7$, at least $10$ times higher than
  what is measured \cite {Lifshitz2000}.

\bibitem{Note2}
This statement is even more true in our experiment than in \cite {Geitner2016},
  since the pressure is lowered by a factor $\sim 10^4$.

\bibitem{Komori2018}
K.~Komori, Y.~Enomoto, H.~Takeda, Y.~Michimura, K.~Somiya, M.~Ando, and S.~W.
  Ballmer, ``Direct approach for the fluctuation-dissipation theorem under
  nonequilibrium steady-state conditions,'' {\em Phys. Rev. D}, vol.~{97},
  p.~102001, May 2018.

\bibitem{Cleland2002}
A.~N. Cleland and M.~L. Roukes, ``Noise processes in nanomechanical
  resonators,'' {\em Journal of Applied Physics}, vol.~{92}, no.~5,
  pp.~2758--2769, 2002.

\bibitem{Akutsu2019}
T.~A. et~al., ``First cryogenic test operation of underground km-scale
  gravitational-wave observatory {KAGRA},'' {\em Classical and Quantum
  Gravity}, vol.~{36}, p.~165008, jul 2019.

\bibitem{Horowitz2020}
J.~M. Horowitz and T.~R. Gingrich, ``Thermodynamic uncertainty relations
  constrain non-equilibrium fluctuations,'' {\em Nature Physics}, vol.~16,
  no.~1, pp.~15--20, 2020.

\bibitem{Schaffer2005}
T.~E. Sch{\"a}ffer, ``Calculation of thermal noise in an atomic force
  microscope with a finite optical spot size,'' {\em Nanotechnology},
  vol.~{16}, pp.~664--670, mar 2005.

\bibitem{Welch1967}
P.~Welch, ``The use of fast fourier transform for the estimation of power
  spectra: A method based on time averaging over short, modified
  periodograms,'' {\em IEEE Transactions on Audio and Electroacoustics},
  vol.~{15}, pp.~70--73, June 1967.

\bibitem{Blackman1958}
R.~Blackman and J.~Tukey, {\em The Measurement of Power Spectra: From the Point
  of View of Communications Engineering}.
\newblock Dover Books on Engineering and Engineering Physics, Dover, 1958.

\bibitem{Nowick1972}
A.~S. Nowick and B.~S. Berry, {\em Anelastic Relaxation in Crystalline Solids}.
\newblock Academic Press, 1972.

\bibitem{Landau1970}
L.~Landau and E.~Lifshitz, {\em Theory of Elasticity}.
\newblock No.~v. 7 in Course of theoretical physics, Elsevier Science, 1970.

\bibitem{Pottier2017}
B.~Pottier and L.~Bellon, ````noiseless'' thermal noise measurement of atomic
  force microscopy cantilevers,'' {\em Applied Physics Letters}, vol.~{110},
  no.~9, p.~094105, 2017.

\bibitem{AOM}
aaptoelectronic.

\bibitem{Metzger2008}
C.~Metzger, M.~Ludwig, C.~Neuenhahn, A.~Ortlieb, I.~Favero, K.~Karrai, and
  F.~Marquardt, ``Self-induced oscillations in an optomechanical system driven
  by bolometric backaction,'' {\em Physical review letters}, vol.~101,
  p.~133903, 10 2008.

\bibitem{Lifshitz2000}
R.~Lifshitz and M.~Roukes, ``Thermoelastic damping in micro- and nanomechanical
  systems. phys. rev. b 61(8), 5600-5609,'' {\em Physical Review B}, vol.~{61},
  pp.~5600--5609, 02 2000.

\end{thebibliography}

\end{document}